\documentclass[12pt]{article}
\usepackage{sc3conf}
\usepackage{amsfonts}
\usepackage{epsfig}
\usepackage{graphicx}

\def\tom{\tilde{\omega}}
\def\toml{\tilde{\omega}_{\rm L}}
\def\omf{\Omega_{\rm F}}
\def\omh{\Omega_{\rm H}}
\def\rl{R_{\rm L}}

\def\xa{x_{\rm A}}

\def\css{c_{\rm s\star}}

\def\sigs{\sigma_{\star}}

\def\up{u_{\rm p}}
\def\vp{v_{\rm p}}
\def\bp{B_{\rm P}}

\def\tbp{\tilde{B}_{\rm p}}

\begin{document}
\raggedbottom

\title{Stationary models of relativistic magnetohydrodynamic jets}

\authors{Christian Fendt}

\addresses{    Institut f\"ur Physik, Universit\"at Potsdam,
               Am Neuen Palais 10,
               14469 Potsdam, Germany
                  }

\maketitle

\begin{abstract}
Highly relativistic jets are most probably driven by strong magnetic
fields
and are launched from the accretion disk surrounding a central black 
hole.

\noindent
In this paper we review some of our recent results considering the 
two-dimensional magnetic field structure and the dynamics of
collimating relativistic jets.
Applying the stationarity assumption enables us to calculate
{\em global} solutions of the relevant MHD equations.
The structure of the jet magnetosphere follows from the so called
Grad-Shafranov equation for the force-balance between axisymmetric
magnetic surfaces. 
The plasma dynamics along the magnetic field lines is given by
the solution of the MHD wind equation.
In the force-free assumption, which is appropriate for relativistic 
jets, both equations de-couple.

\noindent
We discuss solutions of the Grad-Shafranov equation obtained in a 
general relativistic context applying the 3+1 formalism for Kerr
geometry.
These solutions extend from the inner light surface around the Kerr
black hole to the asymptotic regime of a cylindrically collimated
jet with a finite radius.

\noindent
In a further step, we include differential rotation of the foot
points of the field lines.
In this case, the shape of the light surface is not known a priori and
must be calculated in an iterative way.
The solutions show that differentially rotating jets are 
collimated to smaller radii compared to jets with rigid rotation. 

\noindent
Considering the general relativistic MHD wind equation, we investigate
the dynamics of the collimating jet, in particular the effects of Kerr
metric on the acceleration.
Temperature and density follow a power law versus radius.
The jet magnetic (velocity) field is dominated by the toroidal 
(poloidal) component.

\noindent
Having at hand a relativistic MHD jet solution, we calculate the
thermal optically thin X-ray spectrum for the innermost hot part 
of the jet.
Doppler shift and boosting is taken into account.
For microquasars we obtain a jet X-ray luminosity 
$ \approx 10^{33}\,{\rm erg\,s}^{-1}$.
Iron emission lines are clearly visible.

\end{abstract}

\section{Relativistic astrophysical jets}
Relativistic jets originating in the close environment of a rotating 
black hole are observationally indicated for two categories of sources 
concerning mass and energy output.
In active galactic nuclei (AGN), the jets are launched in the magnetized
environment around a rotating, super massive black hole 
(Blandford \& Payne~1982, Sanders et al.~1989).
The discovery of Galactic microquasars has proven superluminal motion 
on a much smaller energy scale (Mirabel \& Rodriguez 1994),
the central engine, however, is believed to work similarly to the AGN.

From the observations we know that astrophysical jets are collimated 
to almost cylindrical shape.
Relativistic jets are generally detected in non-thermal radio
emission, clearly indicating the magnetic character of jet 
formation -- magnetic acceleration and collimation.
There is clear evidence that jet formation is always connected to
the existence of an accretion disk.
As the central object in AGN and microquasars is a black hole, 
the surrounding accretion disk 
is the only possible location for a magnetic field field generation.
Considering the before mentioned observational constraints, it is clear 
that a theoretical, quantitative analysis of the jet structure 
in these sources must take into account both magnetohydrodynamic (MHD)
effects and (general) relativity.
Relativistic MHD implies that electric fields,
which are unimportant in Newtonian MHD, cannot be neglected.

Due to the complexity of the MHD equations, 
all stationary, relativistic models of magnetic jets so far, 
rely on simplifying assumptions
such as self-similarity (Contopoulos 1994), some other
prescription of the field structure 
(Li 1993, Beskin 1997) or the restriction to
asymptotic regimes (e.g. Appl \& Camenzind 1993; 
Nitta 1997; Beskin \& Malyshkin 2000).
As relativistic jets must be highly magnetized, 
the force-free limit of MHD may be applied.
This allows for a truly {\em two-dimensional} (2D) calculation of the
field structure
(Camenzind 1987; Fendt et al.~1995; Fendt 1997; Fendt \& Memola 2001), 
in contrast to the self-similar treatment.

The stationarity assumption itself is essential in order to obtain 
a {\em global} solution of the equations.
Yet, time-dependent MHD simulations of relativistic jet formation are
limited in spatial scale and to a short time evolution
(Koide et al.~2000).
On the other hand, we do not really know whether a stationary MHD 
solution obtained is actually stable. 

\section{Stationary relativistic MHD}
The axisymmetric, stationary electromagnetic equations in the context
of Kerr metric were first derived by Znajek (1977). 
The more comprehensible formalism of a ``3+1 split'' has been
introduced by Thorne et al. (see Thorne et al.~1986).
Camenzind (1986, 1987) formulated a fully relativistic stationary
description of axisymmetric MHD flows presenting solutions
of the Grad-Shafranov and the wind equation.
An essential step to obtain the {\em global} magnetic field structure 
is to take into account properly the regularity condition along the 
light surface.
This matching problem of relativistic magnetospheres has been solved
by Fendt et al.~(1995).
The general relativistic version of the Grad-Shafranov equation
including inertial terms and entropy has been presented 
by Beskin \& Pariev (1993). 
Solutions of the wind equation in Kerr geometry considering the 
stationary plasma motion along the magnetic field
were obtained by Takahashi et al.~(1990), however, 
mainly discussing the accretion flow onto the black hole.

\subsection{Space-time around rotating black holes}
%
In the 3+1 split the space-time around rotating black holes 
(mass $M$, angular momentum per unit mass $a=J/Mc$) can be 
described using the Boyer-Lindquist line element\footnote{
The parameters of the metric tensor are defined as usual,
$\rho^2   \equiv r^2 + a^2\,\cos^2\theta\,,  \quad 
\Delta  \equiv  r^2 - 2\,G\,M\,r /c^2 + a^2\,,$\quad  
$\Sigma^2 \equiv (r^2 + a^2)^2 -a^2\Delta\sin^2\theta\,,\quad 
\tom  \equiv (\Sigma/\rho)\,\sin\theta\,,$ \quad 
$\omega \equiv 2\,a\,G\,M\,r / c\,\Sigma^2\,, \quad
\alpha  \equiv  \rho\,{\sqrt{\Delta}} / \Sigma $.
Here, $\omega $ is the angular velocity of an observer moving with zero
angular momentum (ZAMO), $\omega = (d\phi/dt)_{\rm ZAMO}$
and 
$\alpha $ the lapse function, describing the lapse of proper time $\tau$ 
in the ZAMO system to the global time $t$,
$\alpha = (d\tau/dt)_{\rm ZAMO}$.
}
\begin{equation} 
ds^2 = \alpha^2c^2dt^2 - \tom^2\,(d\phi -\omega dt)^2
-(\rho^2/\Delta)\,dr^2 - \rho^2\,d\theta^2 \,.
\end{equation}
$t$ denotes a global time in which the system is stationary, $\phi$ is
the angle around the symmetry axis, and $r,\theta$ are similar to
the flat space spherical coordinates.
The electromagnetic field $\vec{B}, \vec{E}$, the current density $\vec{j}$,
and the electric charge density $\rho_{\rm c}$ are measured by the ZAMO
according to the locally flat Minkowski space.
These local experiments are put together by a global observer at a 
certain global time using the lapse and shift function to transform from 
local to global frames.
In spite of this transformation, Maxwell's equations in the 3+1 split look
similar to those in Minkowski space. 

\subsection{The cross--field force--balance}
%
The axisymmetric field structure follows from the force-balance across
magnetic flux surfaces. 
%
The magnetic flux function 
$
\Psi (r,\theta) = \frac {1}{2 \pi} \int {\vec {B}}_{\rm P} \cdot d{\vec{A}}
$
measures the magnetic flux
through a loop of the Killing vector $\vec{m} = \tom^2\nabla\phi$.
Similarly, the total poloidal current is defined by integrating 
the poloidal current density over the same loop
$
I = -\int \alpha {\vec {j}}_{\rm P} \cdot d{\vec{A}} 
  = - \frac{c}{2}\alpha\tom B_{\rm T}\,.
$
The indices P and T denote the poloidal and toroidal components of a vector.
In the force-free limit,
$
\rho_{\rm c} \vec{E} + \frac {1}{c} \vec{j}\times\vec{B} = 0\,,
$
we have $\vec{j}_{\rm P} \parallel \vec{B}_{\rm P}$
and, thus, $I = I(\Psi)$.
For a degenerated magnetosphere, 
$
| B^2 - E^2 | >> | \vec{E} \cdot \vec{B} | \simeq 0
$
another conserved quantity can be derived.
The derivative of the time component of the vector potential
defines the iso-rotation parameter
$
 \omf = \omf(\Psi) = - c (dA_0/d\Psi),
$
which is sometimes interpreted as angular velocity of the magnetic 
field lines.

The force-balance across the field can be derived from the
toroidal component of the stationary Amp\`ere's law and is
commonly known as {\em Grad-Shafranov} equation (GSE) 
for the flux function $\Psi(r,\theta)$,
\begin{equation} 
 \tom \nabla \cdot \left( {\alpha\,\frac {D}{\tom^2}}\nabla\Psi\right) 
= \tom\,\frac{\omega-\omf(\Psi)}{\alpha c^2}\omf'(\Psi)\,|\nabla\Psi|^2
- \frac{1}{\alpha\tom}\frac{2}{c^2}\;I(\Psi)\,I'(\Psi)
\end{equation}
(see Okamoto 1992).
Here, 
$
D \equiv 1 - \left({\tom}/{\tom_{\rm L}}\right)^2
$
The ' denotes derivatives $d/d\Psi$,
and $\toml(\Psi)$ the position of the two light surfaces
\footnote{
The `$+$' (`$-$') sign holds for the outer (inner) light surface
with $\omf > \omega$ (resp. $\omf < \omega$)
},
$
\toml^2 = \left(\pm {\alpha \,c}/({\omf - \omega})\right)^2\,.
$

\subsection{Jet acceleration -- the force--balance along the field}
%
The stationary, polytropic, general relativistic MHD flow along 
axisymmetric flux surfaces $\Psi$ can be described by the 
{\em wind equation}\footnote{
with the following abbreviations,
$
k_0 \equiv g_{33} \omf^2 + 2 g_{03} \omf + g_{00},
$
$
k_2 \equiv  1 - \omf (L/E),
$ and 
$
k_4 \equiv
- \left(g_{33} + 2 g_{03} (L/E) + g_{00}(L/E)^2 \right) /
         \left( g_{03}^2 - g_{00} g_{33} \right)
$
}
\begin{equation}
u_{\rm p}^2 + 1  = \left(\frac{E}{\mu}\right)^2
\frac {k_0 k_2  - 2 k_2 M^2 - k_4 M^4}{(k_0 - M^2)^2}
\end{equation}
for the poloidal velocity 
$\up \equiv \gamma \vp/c$
(Camenzind 1986, Takahashi et al.~1990).
The Alfv\'en Mach number $M$ is defined 
via $M^2 = 4 \mu n \up^2 / \tbp^2 $,
with the proper particle density $n$, the specific enthalpy $\mu$,
and the poloidal magnetic field $\tbp = \bp / (g_{00)}+g_{03}\omf) $,
rescaled for mathematical convenience.
The specific total energy density $E(\Psi)$ and total angular
momentum density $L(\Psi)$ are conserved along the flux surfaces.
For a polytropic gas law ($\Gamma \equiv n/m$),
the wind equation can be converted into a polynomial equation
\mbox{
$
\sum_{i=0}^{2n+2m} A_i\,u_{\rm P}^{i/m} = 0
$
}
(Camenzind 1986),
where the coefficients $A_i$ depend on the radius along $\Psi$,
the field strength and the parameters $\omf, E, L, \sigs$.
The magnetization $\sigs(\Psi)$ measures the Poynting flux
in terms of mass flux.

\begin{figure}[tb]
 \begin{center}
\epsfig{file=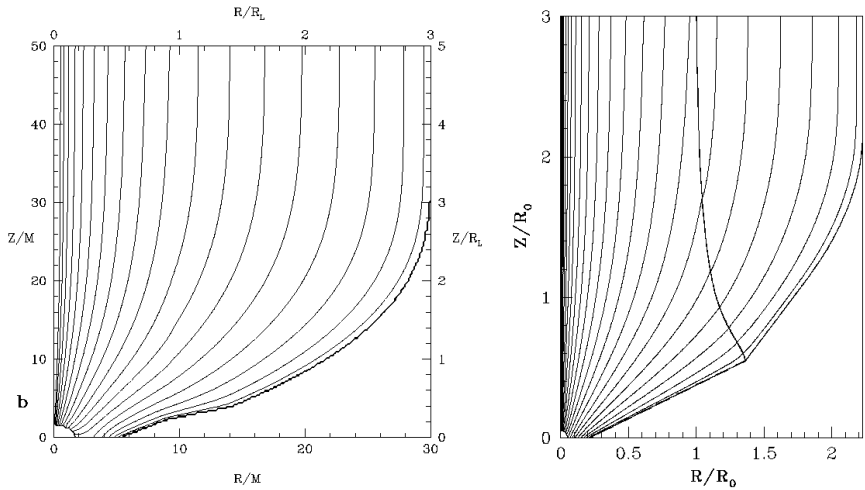, height=6.5cm}
\caption{Axisymmetric magnetic field structure of relativistic jets
$\Psi(R,Z)$.
General relativistic jet magnetosphere around a rotating
black hole with $a=0.8$, $\omf=$const. ({\em left}).
Special relativistic magnetosphere with differential 
rotation $\omf(\Psi)$ ({\em right}).
}
\end{center}
\end{figure}

\section{The 2D structure of relativistic jet magnetospheres}
We now discuss the GSE solution for a force-free, collimating jet
in the general relativistic context (see Fendt 1997),
assuming $\Omega_F(\Psi)= $const.
As asymptotic boundary condition we apply the analytical, 
special relativistic 1D GSE solution
of Appl \& Camenzind (1993).
Since $I=I(\Psi)$, this asymptotic solution provides the GSE 
source term also for the collimation region.
The disk magnetic field distribution is parameterized as 
$\Psi \sim (x/b)^m / (1 + (x/b)^m) + \Psi_H$,
Here, the magnetic flux anchored in the black hole is 
$\Psi_H = 0.5$ with $m=3$ and a core radius $b$.
The asymptotic jet radius can be parameterized in terms of the
light cylinder radius $R_L$ or the gravitational radius $M$.
For $a=0.8$, $(\omf/\omh) =0.4$,
the jet radius is $3\,\rl$ corresponding to $30\,M$.

The main features of the calculated jet magnetosphere are the
following (Fig.\,1, {\it left}).
The field lines originate near the inner light surface close to the
rotating black hole and collimate to an asymptotic jet of finite radius
of several (asymptotic) light cylinder radii.
The solution is defined on a global scale, satisfying the regularity
condition along the light surfaces.
We find a rapid field collimation within $20\,M$ 
distance from the source.
The near-disk solution has three different regimes.
Here, the magnetic flux is either outgoing towards the asymptotic jet
or in-going towards the black hole.
But there exist also flux surfaces near the axis which not connected to 
the disk.
Thus, if we imagine a mass flow associated with the field lines,
we expect three different flow regimes -- accretion, outflow, and a
region empty of a plasma flow.

Extending our work on relativistic magnetospheres, we have 
investigated jets with non-constant $\omf(\Psi)$
(see Fendt \& Memola 2001).
Here, {\em general} relativity is not taken into account.
Since astrophysical jets seem to be always connected to an
accretion disk,
differential rotation of the field lines should be a natural
ingredient for any jet model.
However, 
as a difficulty, then the position and shape of the light
surface singularity is not known a priori, but
have to be calculated iteratively with the field
distribution.
Similar to $I(\Psi)$, also the rotation law 
$\omf(\Psi)$ 
can be taken from the asymptotic GSE solution.

The calculated magnetic field structure is shown in Fig.\,1 
({\it right}).
The differentially rotating magnetosphere is collimated into
a narrower jet.
This, however, can be balanced by an increase of electric current.
Comparison to high-resolution radio observations of the M87 jet
(Junor et al.~1999) shows good agreement in the collimation distance.
For M87 we derive a light cylinder (jet) radius of 50 (120)
Schwarzschild radii.

The field structure is governed by $I(\Psi)$ and $\omf(\Psi)$.
In combination with the disk magnetic flux distribution this
allows to determine 
the magnetic angular momentum loss from the disk in the jet and
the toroidal magnetic field along the disk.
The angular momentum flux per unit time per unit radius is
$d\dot{J}/dx = - x B_{\rm z} I(x)$ along the disk.
In our solutions, most of the magnetic angular momentum is lost
in the outer part of the disk (Fendt \& Memola 2001).

\subsection{The MHD wind solution in Kerr metric} 
Here, we discuss solutions of the stationary, general relativistic, MHD
wind equation (3) along collimating magnetic flux surfaces.
The solutions are calculated on a global scale up to several 1000 
gravitational radii.
Different magnetic field geometries were investigated, parameterized by
the shape of the magnetic flux surface and the flux distribution
(Fendt \& Greiner 2001, FG01).

\begin{figure}[tb]
\begin{center}
\epsfig{file=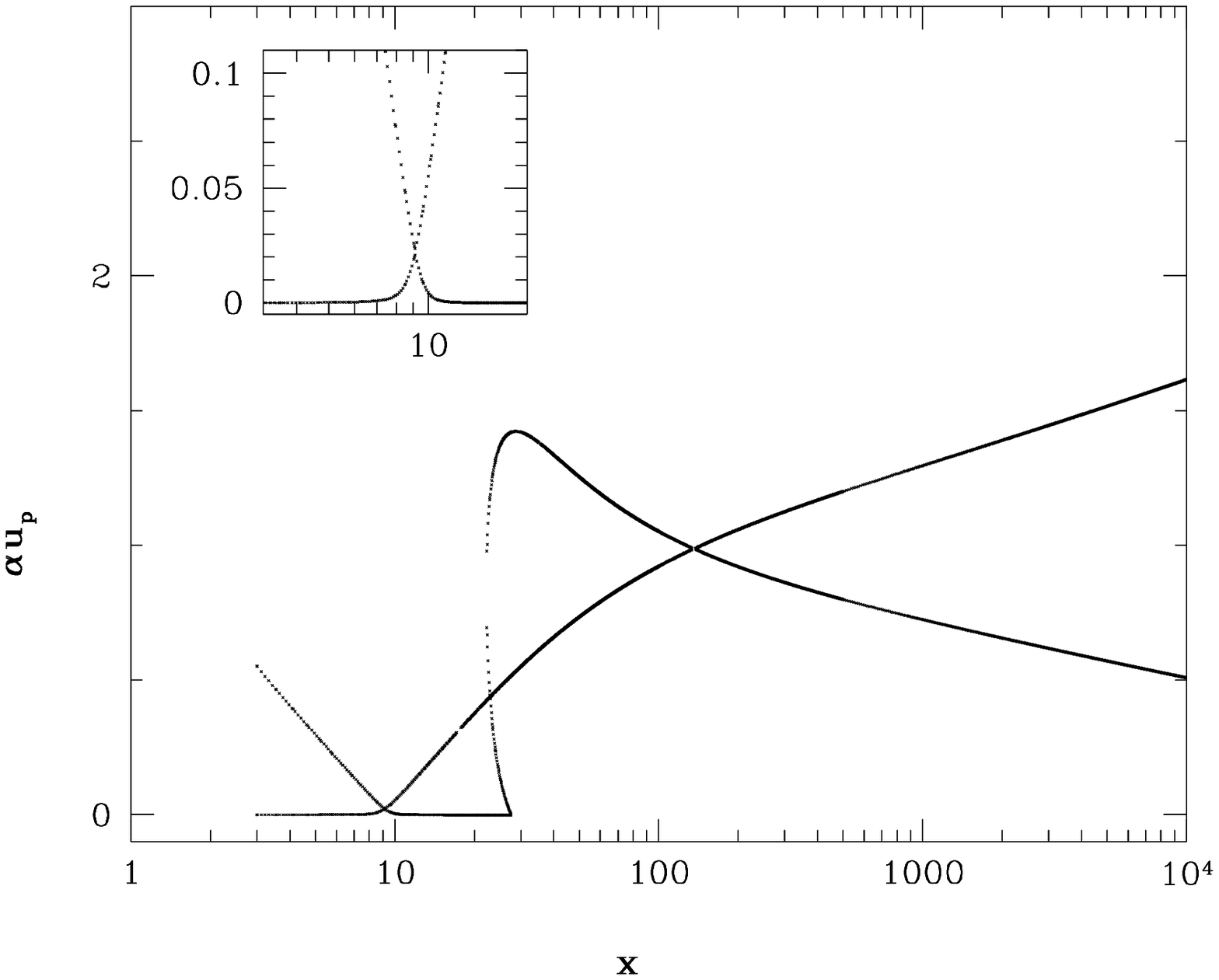, height=5.5cm}
\epsfig{file=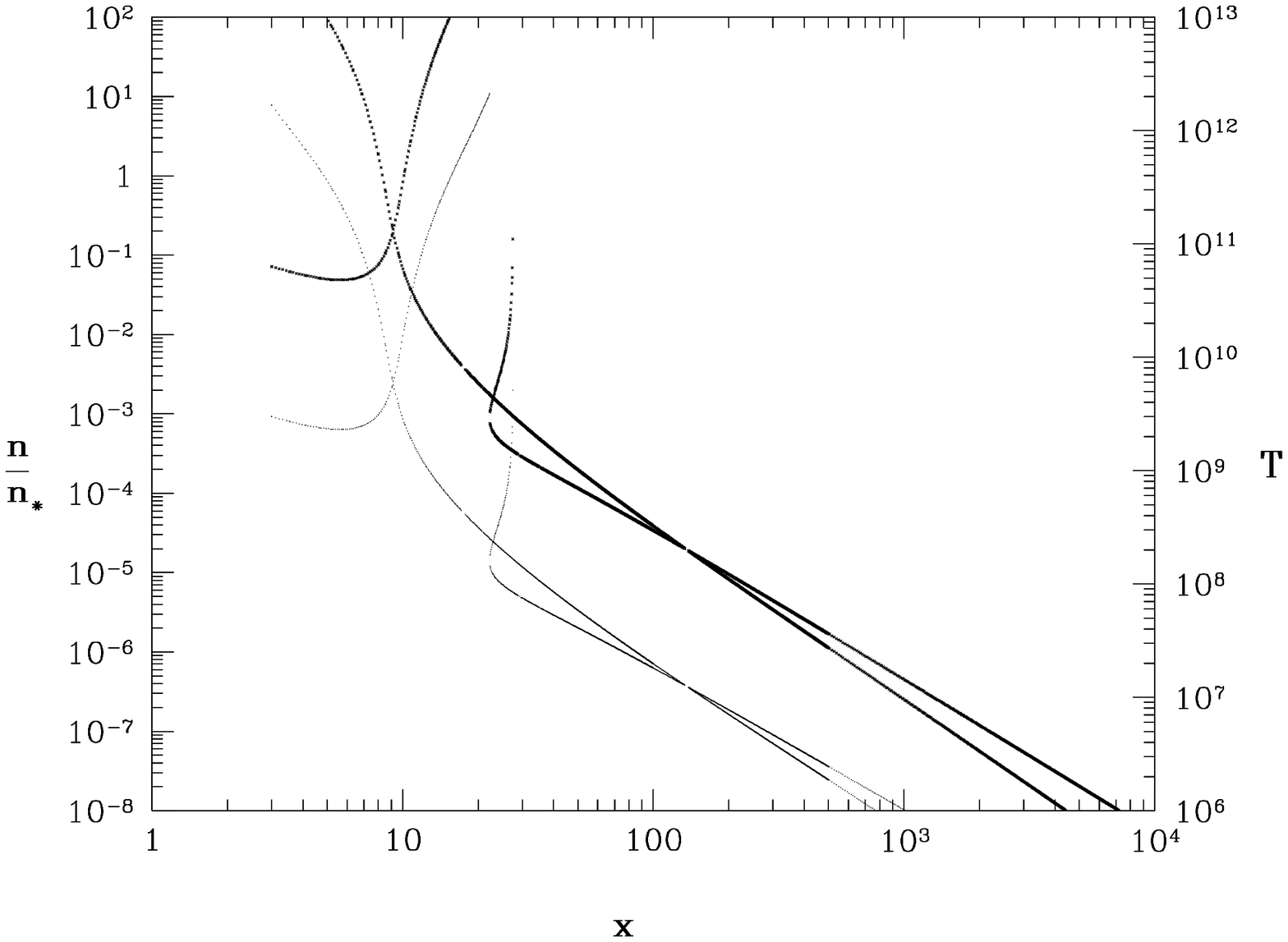, height=5.5cm}
\caption{MHD wind solution in Kerr metric ($a=0.8$). 
Shown is the normalized poloidal velocity $\up$ in the global frame ({\em left})
and the density ({\em thick} line)
and temperature versus radius $x=r/M$ ({\em right}).
The intersections of the physical wind branch with the secondary solution
indicate the magnetosonic points.
Dynamical parameters are (see FG01):
$\sigs = 980 $,  $\omf = 0.035 $,
injection sound speed $\css = 0.05$,
Alfv\'en radius $\xa =  22.9 $,
asymptotic velocity $u_{{\rm p}\infty} = 2.53 $,
total energy $E/m_pc^2 = 2.79 $,
field distribution $z \sim x^{6/5}$ and 
$\tbp \sim x^{-2}$.
}
\end{center}
\end{figure}

Applying microquasar parameters, we obtain the following results (Fig.\,2).
The solution passes all three critical points (slow, Alfv\'en and fast
magnetosonic point).
Substantial acceleration is achieved also beyond the Alfv\'en point due to
the magnetic nozzle effect (Camenzind 1986, Li 1993).
The temperature in the inner jet is up to $10^{11}$K.
The jet injection radius is at $8.3\,M$.
Comparison to solutions with smaller injection radius ($3.3\,M$)
allows to clarify the role of general relativity for the jet
acceleration.
In this case, the asymptotic velocity is much higher due to
the faster rotation $\omf $ at smaller radii.
With $\omf =0.14$ the asymptotic velocity increases to $\up = 8.48$
(FG01).

In the limit of Minkowski metric the jet magnetization can be low, 
although the asymptotic speed is the same as for Schwarzschild metric
(FG01).
The jet flow is not affected by gravity and, 
thus, needs less magnetic energy to gain the same asymptotic
speed by magnetic acceleration.
Solutions for different Kerr parameter $a$ show that the jet is faster
for smaller $a$ (see FG01).
The reason for this seems to be the fact that the effective potential 
of a black hole weakens for increasing values of $a$.

These results might not be surprising as just demonstrating the fact 
of a {\em magnetically driven} jet.
MHD theory tells us that magnetic acceleration takes place mainly
around the Alfv\'en point.
If that is far out, the influence of general relativity must be 
marginal.

\begin{figure}[tb]
\begin{center}
\epsfig{file=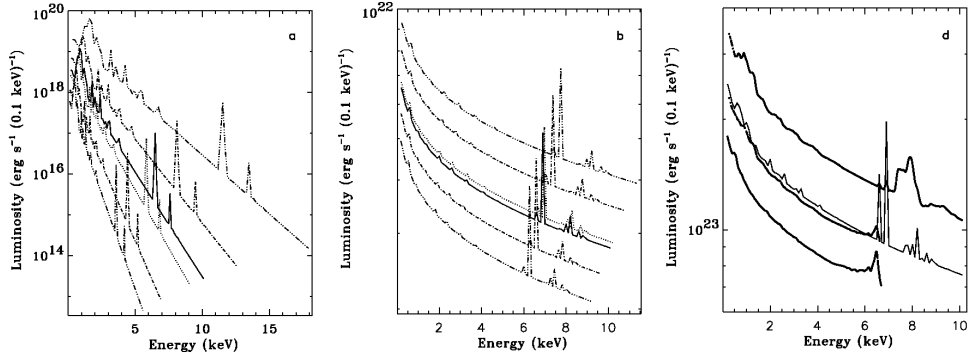, height=4.9cm}
\caption{X-ray spectrum of a relativistic MHD wind solution for a microquasar.
Shown is the luminosity of two volume elements along the jet with different
temperature, $T = 10^7$\,K ({\em left}) and $T = 10^9$\,K ({\em middle}), and
velocity for different inclination,
$i= 40^o, 20^o, 0^o, -20^o, -40^o$ ({\it dashed} curves from
{\it top} to {\it bottom}),
and in the rest-frame ({\it thick} curve).
The {\it right} panel shows the overall spectrum of the red (blue) shifted
side of the jet, $i=-20^o$ ({\it top}) and $i=-20^o$ ({\it bottom}),
the combined spectrum (thick line {\it middle}) and 
the rest-frame spectrum (thin line {\it middle}).
}
\end{center}
\end{figure}

\subsection{X-rays as tracer for jet motion in microquasars?}

The MHD wind solutions discussed above provide the flow dynamics along a
prescribed poloidal magnetic field line with maximum temperatures up to
$10^{11}$\,K in the innermost part of the jet.
Here, we calculate the thermal optically thin X-ray spectrum of such a jet.
Prescribing the jet mass flow rate $\dot{M}_{\rm j}$ together with
the shape of the field line,
the wind solution gives a unique set of parameters of the flow.
For each volume element of the expanding jet flow with decreasing
temperature and density and increasing velocity
we compute the continuum spectrum and the emission lines
of an optically thin plasma (see Memola et al.~2002).
The spectra of the single volumes are then put together to an overall 
spectrum taking into account the Doppler factor 
$
D = (\gamma(1 - \beta \cos\theta))^{-1},
$
for shifting ($E_o = D\,E_e$) and boosting ($L_o(E_o) = D^3 \, L_e(E_e)$)
the emitted energy $E_e$ and luminosity $L_e$.
These factors vary due to the different velocity $\beta$ of the matter in these 
volumes and the different inclination $\theta$ of the velocity vectors to the
line of sight.

Figure 3 shows the spectra of two example volume elements of different
temperature (and velocity) for different jet inclinations.
The combined total spectrum of all volume elements within the collimating
jet cone is shown in Fig.\,3 ({\em right}).
We find a X-ray luminosity $\sim 10^{33}$ 
($\dot{M}_{\rm j}/10^{-8}\,{\rm M}_{\odot}{\rm yr}^{-1}){\rm erg\,s^{-1}}$
for a microquasar.
The 6.6 and 6.9\,keV emission lines can be identified
as $K\alpha$ iron lines, while the one
at 8.2\,keV could be the $K\beta$.
We emphasize that our approach is not a {\em fit} to
certain observed spectra.
In contrary, for the first time, for a jet flow with characteristics defined
by the solution of the MHD wind equation, we derive its X-ray spectrum.

\end{document}